\documentclass[superscriptaddress,nofootinbib,letterpaper,prb,twocolumn]{revtex4-2}

\usepackage{amsmath,amssymb,amsfonts}
\usepackage{mathrsfs}
\usepackage{graphicx}
\usepackage[usenames,dvipsnames]{color}
\usepackage{hyperref}
\usepackage{url}
\usepackage[utf8]{inputenc}
\usepackage{slashed}
\usepackage{subfigure}
\usepackage{slashed,bbm}
\usepackage{graphics,psfrag,epsfig}
\usepackage{dsfont}
\usepackage{setspace}
\usepackage{bm}

\makeatletter
\newcommand*{\toccontents}{\@starttoc{toc}}
\makeatother

\usepackage{hyperref}
\usepackage{xcolor}
\definecolor{dark-red}{rgb}{0.4,0.15,0.15}
\definecolor{dark-blue}{rgb}{0.15,0.15,0.4}
\definecolor{medium-blue}{rgb}{0,0,0.5}

\allowdisplaybreaks

\begin{document}

\title{$\mathbf{\mathcal{N}=4}$ chiral superconductivity in moir\'e transition metal dichalcogenides}

\author{Michael M. Scherer}
\affiliation{Institut f\"ur Theoretische Physik, Universit\"at zu K\"oln, 50937 Cologne, Germany}

\author{Dante M. Kennes}
\affiliation{Institute for Theory of Statistical Physics, RWTH Aachen University,
and JARA Fundamentals of Future Information Technology, 52056 Aachen, Germany}
\affiliation{Max Planck Institute for the Structure and Dynamics of Matter,
Center for Free Electron Laser Science, Luruper Chaussee 149, 22761 Hamburg, Germany}

\author{Laura Classen}
\affiliation{Condensed Matter Physics \& Materials Science Division, Brookhaven National Laboratory, Upton, NY 11973-5000, USA}

\maketitle


\textbf{Experimental demonstrations of tunable correlation effects in magic-angle twisted bilayer graphene~\cite{Cao2018a, Cao2018, Yankowitz2019,Kerelsky2019, Sharpe605, lu2019superconductors, Serlin2019} have put two-dimensional moir\'e quantum materials at the forefront of condensed-matter research~\cite{Kennes2021}.
Other twisted few-layer graphitic structures~\cite{Liu2019, Cao2019, Shen2019,Chen2019a, Chen2019, chen2019tunable,TutucBi,rubioverdu2020universal}, boron-nitride~\cite{Xian2019BN}, and homo- or hetero-stacks of transition metal dichalcogenides (TMDs)~\cite{Wu2018, Wu2019, Naik2018, Ruiz-Tijerina2019,Wu2018, Wu2019, Schrade2019,Wang2020WSe2,zhou2021quantum} have further enriched the opportunities for analysis and utilization of correlations in these systems.
Recently, within the latter material class, strong spin-orbit coupling~\cite{Wang2020WSe2,zhou2021quantum} or excitonic physics~\cite{Jin2019ex,Wang2019ex,shimazaki2020stronglyex} were experimentally explored.
The observation of a Mott insulating state~\cite{tang2019wse2,Regan2020} and other fascinating collective phenomena such as generalized Wigner crystals~\cite{Regan2020}, stripe phases~\cite{Jin2021} and quantum anomalous Hall insulators~\cite{li2021quantum} confirmed the relevance of many-body interactions, and demonstrated the importance of their extended range.
Since the interaction, its range, and the filling can be tuned experimentally by twist angle, substrate engineering and gating, we here explore Fermi surface instabilities and resulting phases of matter of hetero-bilayer TMDs.  
Using an unbiased renormalization group approach, we establish in particular that hetero-bilayer TMDs are unique platforms to realize topological superconductivity with winding number~$|\mathcal{N}|=4$. 
We show that this state reflects in pronounced experimental signatures, such as distinct quantum Hall features.
}

The pairing of electrons in a superconductor is among the most intriguing effects in the study of collective phenomena. 
In the quest to achieve ever higher critical temperatures unconventional superconducting states have received an increasing amount of attention \cite{RevModPhys.75.473,RevModPhys.79.353} as they allow superconducting temperatures beyond the bounds of standard BCS theory \cite{Bardeen1209}.
At the same time, combining superconductivity with non-trivial topology is a promising route for quantum information sciences as such topological superconductors may harbor robust edge states at domain boundaries with topological properties advantageous to computing applications \cite{RevModPhys.80.1083}. 

However, realizing and controlling  topological superconductors proves difficult to this date, with only a few candidate materials currently being suggested, e.g., \cite{UPTRMP,UPt3,UTe2,Zhang182,Li2021}. 
A new direction in the study of superconductivity opened up recently in twisted moir\'e quantum materials, i.e. two-dimensional van der Waals materials being stacked at a relative twist angle~\cite{Cao2018a, Cao2018, Yankowitz2019,Kerelsky2019, Sharpe605, lu2019superconductors,Serlin2019,Kennes2021,Liu2019, Cao2019,Shen2019,Chen2019a,Chen2019,chen2019tunable,TutucBi,rubioverdu2020universal,Xian2019BN,Lian20,KennesGeSe}.
In these systems kinetic energy scales can be tuned by the twist angle allowing to promote the relative relevance of potential, spin-orbit coupling or other energy scales \cite{Kennes2021}.
Indeed, topological properties as well as superconductivity were already demonstrated in these highly versatile systems and as a consequence they could provide an excellent opportunity to engineer novel topological superconductors.

\begin{figure*}[t!]
\begin{center}
\includegraphics[width=\textwidth]{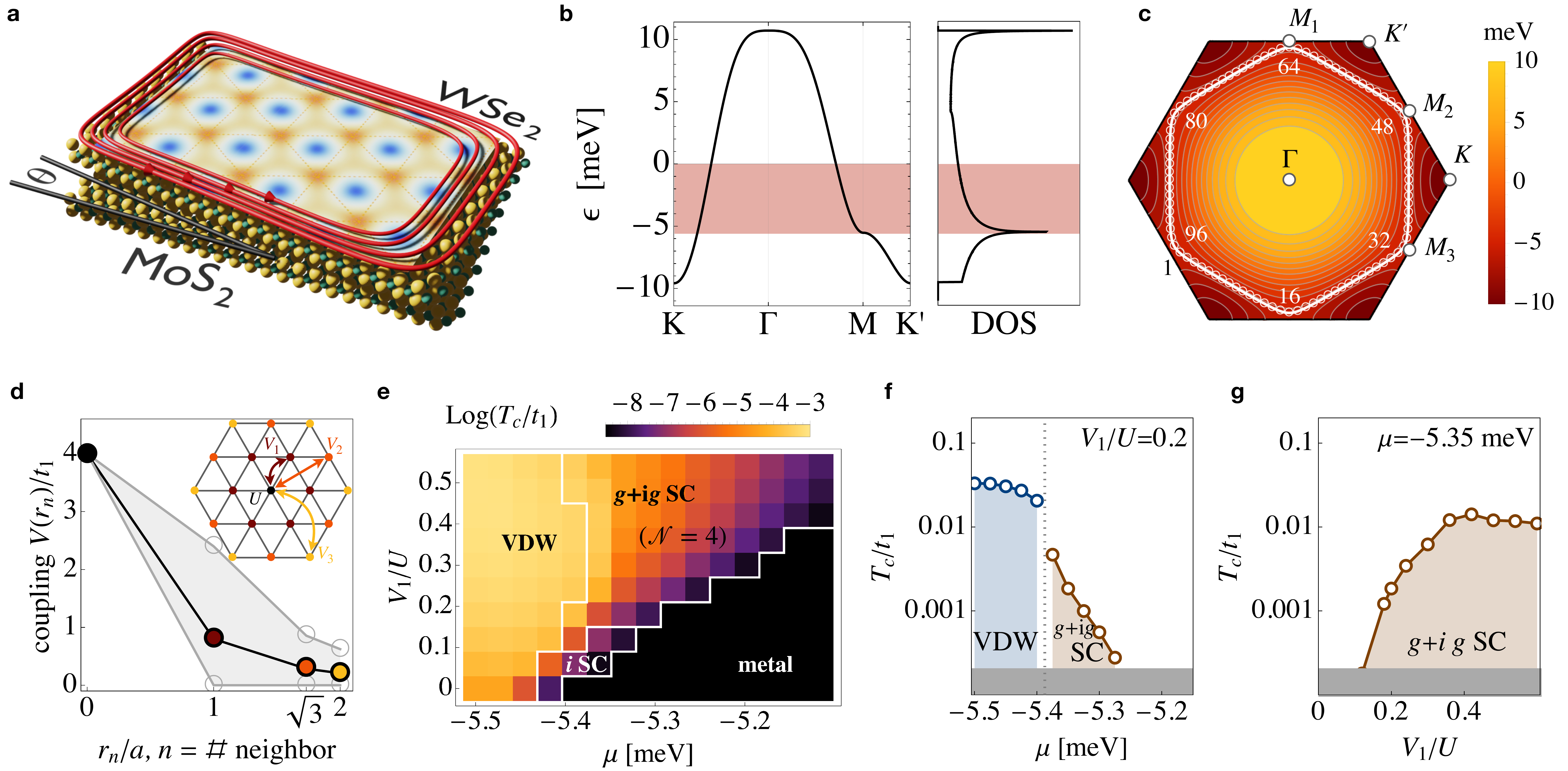}
\end{center}
\caption{\textbf{Correlated phase diagram of hetero-bilayer TMDs.} 
\textbf{a}~Sketch of a twisted WSe$_2$/MoS$_2$ bilayer with small twist angle $\theta$. The resulting effective moir\'e potential \cite{Wu2018,zhou2021quantum} is indicated by the contour shading at the top.
The four red lines represent the chiral edge modes of the topological superconducting state with $\mathcal{N}=4$. \textbf{b}~Dispersion of the energy band $\epsilon_{\vec k}$ along high-symmetry lines and density of states (DOS). 
We explore filling levels indicated by the red band between half filling ($\mu\sim 0$) and Van Hove filling ($\mu=-5.5$meV, $\sim 1/4$ filling).
\textbf{c}~Nearly nested Fermi surface close to the Van Hove energy. 
Numbered open circles show the momentum resolution of the Fermi surface employed in the FRG approach. 
\textbf{d}~ Extended interaction parameters as function of distance $r_n/a$, $a$ is the moir\'e lattice spacing \cite{Wu2018,zhou2021quantum}. The gray area marks the explored range of interactions, which are tunable by the environment. 
Inset: sketch of the interactions on the triangular moir\'e lattice.
\textbf{e}~Phase diagram as function of nearest-neighbor repulsion $V_1$ and filling controlled by the chemical potential $\mu$ extracted from the Fermi-liquid instabilities in the FRG flow for $U=4t_1$, $V_2/V_1=0.357$ and $V_3/V_1=0.260$ (see Fig.~\ref{fig:FS}d). 
The background color encodes our estimate for the critical temperature (see methods).  
We find an instability towards a valley density wave (VDW) near Van Hove filling $\mu\approx -5.5$meV.  
It is flanked by an $i$-wave superconducting instability ($i$SC) for small $V_1/U$ and by a $g$-wave superconducting instability for larger $V_1/U$. 
In the ground state, the $g$-wave superconducting instability forms a chiral $g+ig$ state ($g+ig$ SC) characterized by winding number $\mathcal N=4$. 
The regime colored in black, does not show any instability within our numerical accuracy.
\textbf{f} Filling-dependent critical temperature near Van Hove filling. Below the gray band, we cannot resolve any instability within our numerical accuracy. 
\textbf{g} Dependence of the critical temperature on the extended interactions tuned via $V_1/U$ at fixed $V_2/V_1, V_3/V_1$ slightly away from Van Hove filling.
}
\label{fig:FS}
\end{figure*}

Here, we explore this idea for moir\'e transition metal dichalcogenides (see Fig.~\ref{fig:FS}a) \cite{Wu2018, Wu2019, Naik2018, Ruiz-Tijerina2019,Wu2018, Wu2019, Schrade2019,Wang2020WSe2,zhou2021quantum,Kennes2021,tang2019wse2} by analyzing the Fermi surface instabilities of twisted hetero-bilayers of  WX$_2$/MoX$_2$ (X=S,Se) away from half filling of the moir\'e band.
We unveil an exotic superconducting state near Van Hove filling described by form factors with eight zero crossings, arising from the extended range of interactions in these materials. 
We show that the superconducting ground state is formed by a chiral configuration, which is characterized by a full gap on the Fermi surface and non-trivial topology with winding number $|\mathcal{N}|=4$. 
We argue that this type of topological superconductivity leads to distinct experimental signatures in quantum Hall transport measurements and elevates twisted hetero-bilayers of TMDs to prime candidates for experimental scrutiny of topological superconductivity. 

\begin{figure*}[t!]
\begin{center}
\includegraphics[width=\textwidth]{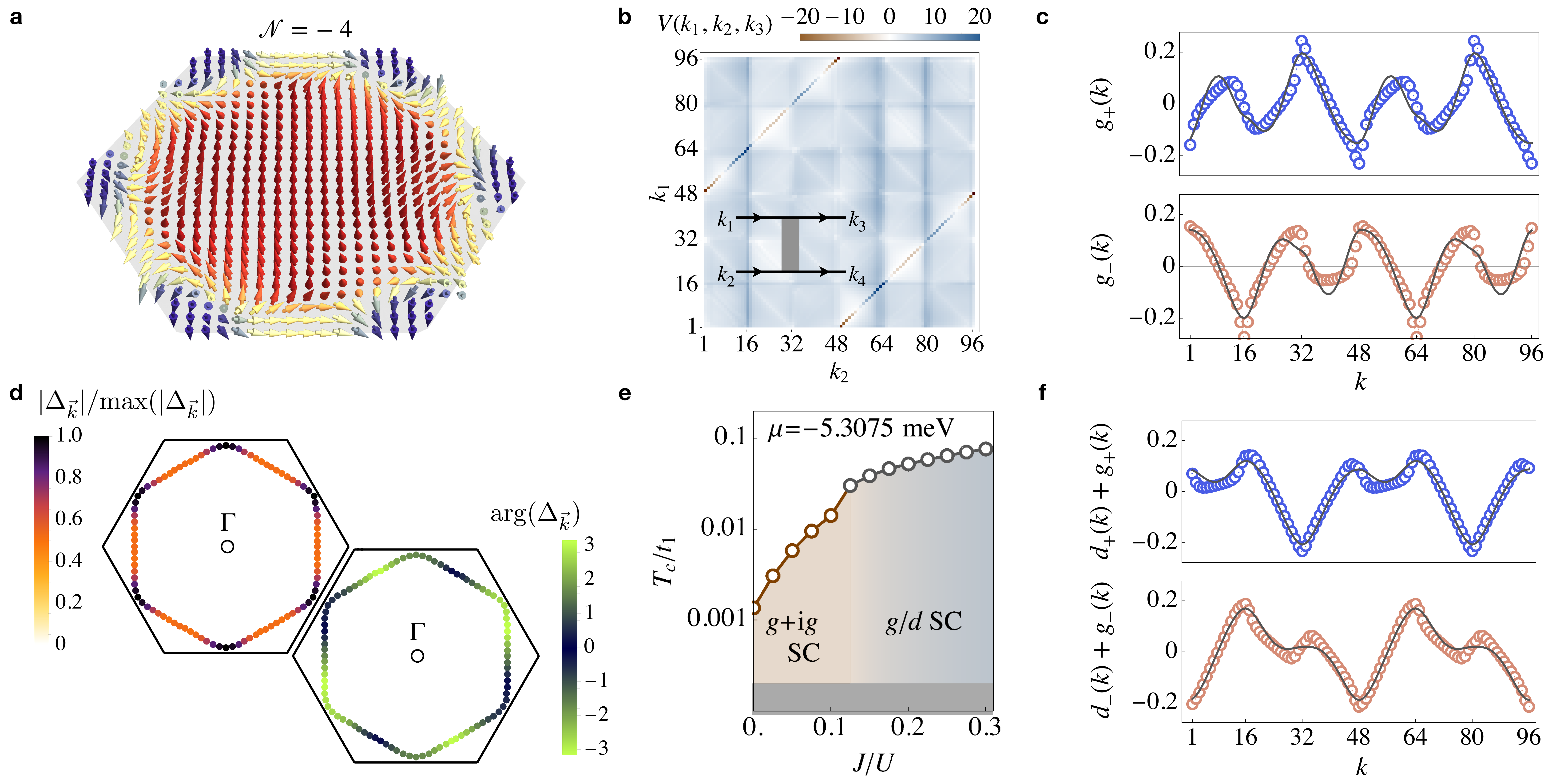}
\end{center}
\caption{\textbf{Topological superconductivity in hetero-bilayer TMDs.}
\textbf{a}~The phase winding of the superconducting gap in the $g+ig$ state can be visualized by a skyrmion configuration constructed from the vector $\vec{m}=(\mathrm{Re}\Delta_{\vec{k}},\mathrm{Im}\Delta_{\vec{k}},\xi_{\vec{k}})/(\xi_{\vec k}^2+|\Delta_{\vec k}|^2)^{1/2}$. 
$\vec m$ points up (down) at the highest (lowest) energies $\xi_{\vec k}=\epsilon_{\vec k}-\mu$ and rotates $\mathcal N$ times in the plane in the vicinity of the Fermi level $\xi_{\vec k}=0$.
The skyrmion configuration is used to calculate the winding number $\mathcal{N}$, which for a broad range of fillings around Van Hove filling is $|\mathcal{N}|=4$, indicating an enhanced response in thermal and spin quantum Hall measurements.
\textbf{b}~Functional RG data of the irreducible two-particle correlation function $V(\vec k_1, \vec k_2, \vec k_3,  \vec k_4)$ near the instability temperature for incoming wave vectors $\vec k_1,\vec k_2$. 
Wave vectors are labeled by the patch points along the Fermi surface indicated in Fig.~\ref{fig:FS}c.
The outgoing wave vector $\vec k_3$ is fixed at patch no.~1 and $\vec k_4=\vec k_1+\vec k_2-\vec k_3$ is given by momentum conservation.
The sharp diagonal features occur at $\vec k_1=-\vec k_2$, $\vec k_3=-\vec k_4$, indicating the formation of long-ranged superconducting correlations.
\textbf{c}~Superconducting form factors $g_\pm(k)$ extracted from $V(\vec k_1, \vec k_2, \vec k_3,  \vec k_4)$ in b.
They exhibit a large overlap with a linear combination of the second-nearest-neighbor lattice harmonics $g_1(\vec k),\;g_2(\vec k)$ (solid gray lines) defined in the text, which belong to the two-dimensional irreducible representation $E_2$ of the lattice symmetry group $C_{6v}$. 
We classify them as $g$-wave form factors due to their eight nodes.
\textbf{d}~Absolute value and phase of the gap function on the Fermi surface.
The chiral superposition $\Delta_{\vec k}=|\Delta| (g_1(\vec k)\pm i g_2(\vec k))$ fully gaps the Fermi surface, thereby minimizing the energy. 
Such a $g+ig$ superconducting state breaks time-reversal symmetry and is topological with a four-fold phase winding along the Fermi surface $|\mathcal{N}|=4$.
\textbf{e}~Stability of the $g$-wave superconduting state towards inclusion of  $J/U$ for $V_1/U=0.2$ at $\mu=5.3075$~meV.
For growing values of the exchange interaction $J$, the nearest-neighbor harmonics $d_1$ and $d_2$ of $E_2$ (defined in the text) start to contribute as indicated by the colored transition. 
They are pure $d$-wave form factors with only four nodes.  
For $J/U \lesssim 0.1$, the contribution from $d_1$ and $d_2$ is negligible.
\textbf{f}~Example of the extracted form factors for $J/U=0.5$ where $d_1,d_2$ and $g_1,g_2$ roughly contribute by equal amounts, showing the change in the number of nodes due to the admixture.
}
\label{fig:FS2}
\end{figure*}


In a range of small twist angles, isolated and narrow moir\'e bands emerge in TMD hetero-bilayers of WX$_2$/MoX$_2$ (X=S,Se)~\cite{Wu2018, Wu2019, Naik2018, Ruiz-Tijerina2019,Schrade2019,Wang2020WSe2,zhou2021quantum}.
These flat bands are formed by the highest, spin-polarized valence band of WX$_2$ and can be described 
by an extended triangular-lattice Hubbard model $H=H_0+H_I$, which features an effective SU(2) valley symmetry~\cite{Wu2018} 
\begin{align}
H_0 &= \sum_{v=\pm}\Big[\sum_{ij} t_{i-j}c^\dagger_{i,v}c_{j,v} -\mu\sum_i c^\dagger_{i,v}c_{i,v} \Big]\\
H_I &= U\sum_i n_{i,+}n_{i,-} + \sum_{ij} V_{i-j}n_{i}n_{j}\,.
\label{eq:model}
\end{align}
Here, $n_{i}=\sum_v n_{i,v}$ and $n_{i,v}=c_{i,v}^\dagger c_{i,v}$ is the number of electrons on site $i$ with valley index $\pm$,  $c_{i,v}^{(\dagger)}$ are the corresponding annihilation (creation) operators. 
The hopping amplitudes $t_{n}$ for the $n$th-nearest neighbors depend on the twist angle and we consider typical values for vanishingly small twist angle $t_1 \approx 2.5\, \mathrm{meV}, t_2 \approx -0.5\, \mathrm{meV}, t_3 \approx -0.25\, \mathrm{meV}$~\cite{Wu2018}. 
The resulting moir\'e band $\epsilon_{\vec k}$ features a Van Hove peak in the density of states near 1/4 filling ($-5.5$meV), where the Fermi surface is approximately nested (Fig.~\ref{fig:FS}b,c). 
In experiment, the filling can be adjusted, and  Van Hove filling can be reached, by tuning the gate voltage, which we model here by varying the chemical potential $\mu$ between 1/4 and 1/2 filling.
The interaction parameters $U$, $V_n$ also depend on the twist angle, and on the dielectric environment so that the strength and range of interactions can be controlled~\cite{duran2021moire}.
First-principles calculations show that the extended interactions $V_n$ are sizable in effective models for hetero-bilayer TMDs \cite{Wu2018}. 
For our analysis, we use an intermediate interaction strength for the onsite interaction $U=4 t_1$ and explore the effect of further-ranged interactions by varying $V_1/U\in [0,0.5]$ with $V_2/V_1 \approx 0.357$ and $V_3/V_1 \approx 0.260$~\cite{Wu2018,zhou2021quantum} (Fig.~\ref{fig:FS}d). 
In a second step we also investigate the impact of an additional nearest-neighbor exchange interaction $H_J=J\sum_{\langle i,j \rangle}\vec S_i \vec S_j$ to model strong-coupling effects.


We study the correlated phases of hetero-bilayer TMDs that emerge out of a metallic state within an itinerant scenario using the fermionic functional renormalization group (FRG)~\cite{RevModPhys.84.299}. 
The FRG resolves the competition between different ordering tendencies in an unbiased way and is employed to calculate the dressed, irreducible two-particle correlation function $V(\vec k_1,\vec k_2,\vec k_3,\vec k_4)$ for electrons with momenta $\vec k_i$, $i=1\ldots 4$, on the Fermi surface (Fig.~\ref{fig:FS}c). 
Upon lowering the temperature, $V(\vec k_1,\vec k_2,\vec k_3,\vec k_4)$ develops sharp, localized peaks for characteristic momentum combinations, indicating long-ranged correlations in real space. 
This allows us to extract the temperature where a strongly-correlated state forms, as well as the symmetry and type of the strongest correlations (see methods).

In our model for hetero-bilayer TMD  moir\'e systems, instabilities near 1/4 filling occur due to the high density of states and approximate nesting, which leads to symmetry-broken ground states.
We start with varying $\mu$ and $V_n$ and calculate the phase diagram based on the two-particle correlation functions (Fig.~\ref{fig:FS}e).
Closest to Van Hove filling $\mu\approx -5.5$meV, we find that correlations corresponding to a valley density wave (VDW) are strongest, which manifest themselves by peaks at the nesting momenta $\vec Q_\alpha$, $\alpha=1,2,3$ in $V(\vec k_1,\vec k_2, \vec k_3, \vec k_4)$, i.e. at $k_3-k_1=Q_\alpha$ or $k_3-k_2=Q_\alpha$. 
This state is the analogue of a spin density wave~\cite{PhysRevLett.108.227204,PhysRevLett.101.156402} considering that, here, the SU(2) symmetry belongs to a pseudo-spin formed by the valleys.
The VDW instability is insensitive towards the inclusion of $V_n$ in the explored range.

Moving the filling slightly away from Van Hove filling, we obtain a superconducting instability, which is indicated by diagonal peak positions, i.e. $V(\vec k_1,\vec k_2,\vec k_3,\vec k_4)\approx V(k_1,-k_1,k_3,-k_3)$, that correspond to electron pairs with a total momentum of zero $k_1+k_2=k_3+k_4=0$ (Fig.~\ref{fig:FS2}b).
Increasing the filling further reduces the critical temperature until it vanishes (Fig.~\ref{fig:FS}f). 
The inclusion of $V_n$ has a profound impact:
it strongly affects the symmetry of the superconducting correlations, because it penalizes electrons to be simultaneously on neighboring sites, so that electron pairing is shifted to farther-distanced neighbors. 
As a result, the largest attraction is promoted to occur in a higher-harmonic channel.

The symmetry of the pair correlations can be classified in terms of the irreducible representations of the lattice point group $C_{6v}$ by expanding the eigenfunctions of $V(\vec k,-\vec k,\vec k', -\vec k')$ in lattice harmonics. 
Within an irreducible representation, lattice harmonics with the same symmetry but different angular-momentum form factors can mix and it depends on microscopic details which lattice harmonics are the strongest. 
 
For small $V_n$, we find a small regime with $A_2$ symmetry ($i$-wave) in agreement with previous results for $V_n=0$ \cite{PhysRevB.89.144501,PhysRevB.68.104510}. 
However, for larger $V_n$ ($V_1/U\gtrsim0.15$),  including realistic values in twisted TMDs \cite{Wu2018,zhou2021quantum}, we unveil a large regime with a different symmetry.
That $V_1$ drives this instability can also be seen at the critical temperature, which initially increases with $V_1$ and then saturates (see Fig.~\ref{fig:FS}g).
The pair correlations in this regime are fitted well using the second-nearest-neighbor lattice harmonics $g_1(\vec k)=8/9[-\cos(3 k_x/2) \cos(\sqrt{3} k_y/2) +  \cos(\sqrt{3} k_y)]$,  $g_2(\vec k)=8/(3\sqrt{3})\sin(3 k_x/2) \sin(\sqrt{3} k_y/2)$ (Fig.~\ref{fig:FS2}c). 
They belong to the two-dimensional irreducible representation $E_2$, which contains both, $d$- and $g$-wave form factors. While they cannot, in principle, be distinguished by symmetry, we can categorize our result as $g$-wave based on the number of nodes~\cite{note1}.
This has unique, measurable consequences for the topological properties of the superconducting state. 


Due to the two-dimensional $E_2$ symmetry, the superconducting gap has two components $\Delta_1,\Delta_2$ and additional symmetries besides $U(1)$ can be broken depending on the configuration that forms the ground state~\cite{RevModPhys.63.239}. 
The ground-state configuration is determined by minimizing the Landau energy functional
\begin{align}
    \hspace{-0.15cm}\mathcal{L}\!=\!\alpha(|\Delta_1|^2\!\!+\!|\Delta_2|^2)\!+\!\beta(|\Delta_1|^2\!\!+\!|\Delta_2|^2)^2\!+\!\gamma|\Delta_1^2\!+\!\Delta_2^2|^2.
    \label{eq:Landau}
\end{align}
We use our FRG results as an input for the effective interaction $V(\vec k,-\vec k,\vec k',-\vec k')c_{\vec k',v}^\dagger c_{-\vec k',v'}^\dagger c_{-\vec k,v'} c_{k,v}$ close to the instability and perform a Hubbard-Stratonovich decoupling with the pairing fields $\Delta_i\sim g_i(\vec k) c_{\vec k,+} c_{-\vec k,-}$, $\Delta_i^*\sim g_i(\vec k) c_{-\vec k,-}^\dagger c_{\vec k,+}^\dagger$. 
Integrating out the electrons, we find in particular $\gamma>0$. 
Thus, the chiral configuration $\Delta_1= i \Delta_2$ minimizes the energy. 
Such a ``$g+ig$'' superconducting state breaks time-reversal symmetry and is topologically non-trivial. The Fermi surface is fully gapped as we can see from the quasiparticle energy $E_{\vec k}=(\xi_{\vec k}^2+|\Delta_{\vec k}|^2)^{1/2}$, where $\xi_{\vec k}=\epsilon_{\vec{k}}-\mu$,  $\Delta_{\vec k}=\Delta [g_1(\vec k)+i g_2(\vec k)]$, and $g_1,g_2$ are the FRG-extracted form factors (see Figs.~\ref{fig:FS2}c,d). 
The topological properties can be classified by an integer invariant based on the Skyrmion number \cite{Volovik1997,PhysRevB.61.10267,Black_Schaffer_2014}
\begin{align}
 \mathcal{N} = \frac{1}{4\pi}\int_{\mathrm{BZ}}d^2k\, \vec{m}\cdot\left(\frac{\partial \vec{m}}{\partial k_x}\times\frac{\partial \vec{m}}{\partial k_y}\right)\,,
\end{align}
where the pseudo-spin vector is given by $\vec{m}=(\mathrm{Re}\Delta_{\vec{k}},\mathrm{Im}\Delta_{\vec{k}},\xi_{\vec{k}})/E_{\vec k}$ (Fig.~\ref{fig:FS2}a). 
$\vec m$ follows the phase winding of the superconducting gap around the Fermi surface.
We calculate $\mathcal{N}$ for the entire range of fillings and find $|\mathcal N|=4$ in the relevant regime where the superconducting instability occurs.
Importantly, the high winding number $|\mathcal N|=4$ implies stronger experimental signatures compared to other topological superconductors.  
Four chiral edge modes appear (as illustrated in Fig.~\ref{fig:FS}a) and the quantized thermal and spin response is enhanced with the spin Hall conductance given by $\sigma_{xy}^s=\mathcal N \hbar/(8\pi)$ and the thermal Hall conductance by $\kappa=\mathcal N \pi k_B^2/(6\hbar)$~\cite{PhysRevB.60.4245,PhysRevB.68.214503}.

\begin{figure}[t!]
\begin{center}
\includegraphics[width=\columnwidth]{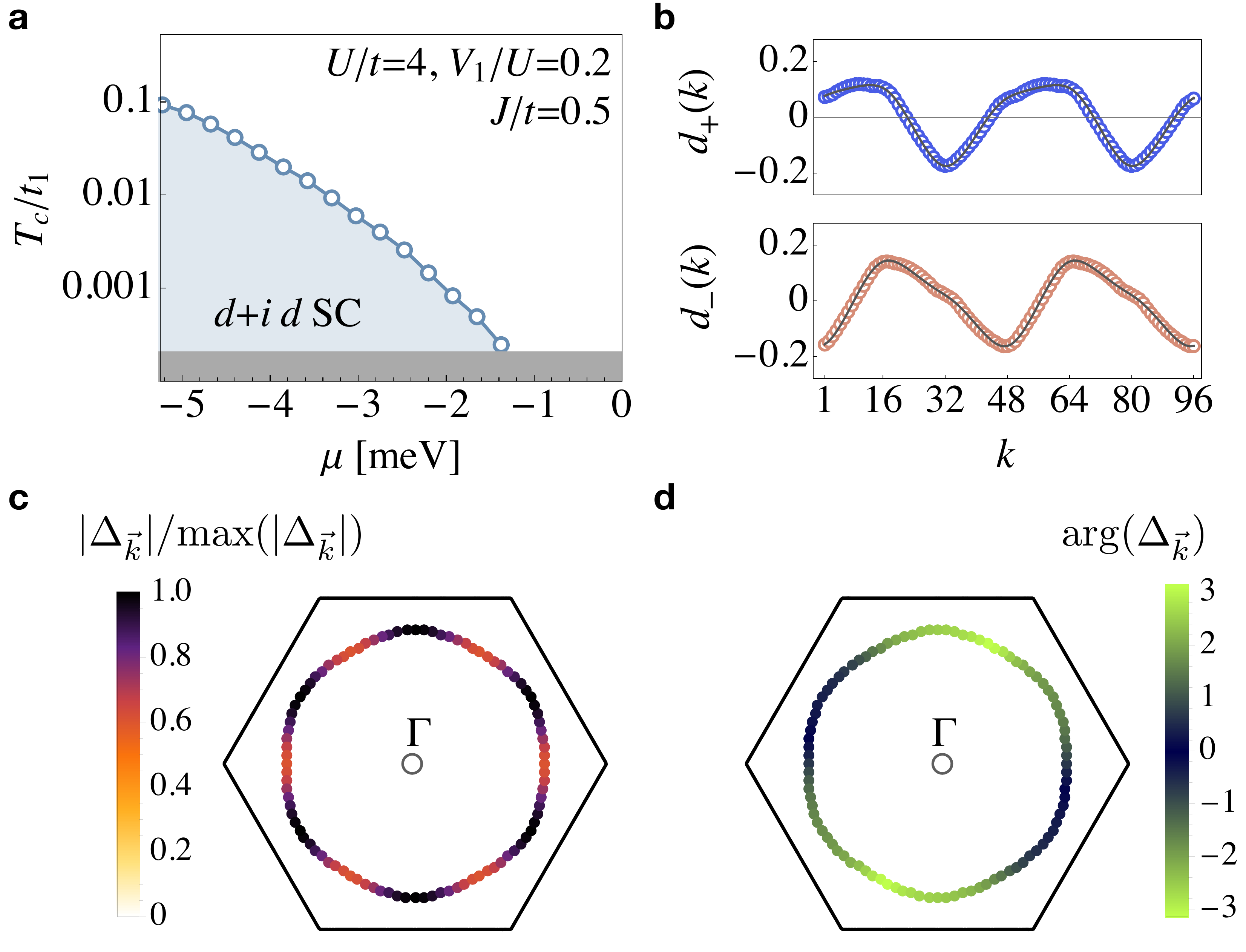}
\end{center}
\caption{\textbf{Effect of exchange coupling.}
\textbf{a} Filling-dependent critical temperature for a sizable exchange coupling $J/t=0.5$. The exchange coupling generates a $d$-wave instability, which indicates a chiral $d+id$ ground state. 
\textbf{b} Superconducting form factors $d_\pm(k)$ extracted from $V(\vec k_1, \vec k_2, \vec k_3,  \vec k_4)$ which exhibit a large overlap with a linear combination of nearest-neighbor lattice harmonics $d_1$, $d_2$ (solid gray lines) defined in the text. 
Their chiral superposition $d\pm id$ fully gaps the Fermi surface with the gap function $\Delta(k)$ shown in the lowest panels (\textbf{c,d}).
The phase winds twice around the Fermi surface. 
}
\label{fig:dwave}
\end{figure}


The $g+ig$ pairing state is robust with respect to the inclusion of small to intermediate nearest-neighbor exchange $J$.
For example, for $V_1/U=0.2$ and $\mu \approx -5.31\,$meV, $g+ig$ pairing is dominant up to reasonably large values of $J \approx 0.1$ (Fig.~\ref{fig:FS2}e).
For larger values of $J$, $d$-wave contributions from the nearest-neighbor form factors $d_1(\vec k)=2 (\cos k_x - \cos (k_x/2) \cos(\sqrt3 k_y/2))$ and $d_2(\vec k)=3/\sqrt{3} \sin (k_x/2) \sin(\sqrt{3} k_y/2)$ start to mix with the previous $g$-wave ones $g_1, g_2$ (see Figs.~\ref{fig:FS2}e,f).
This is expected when the attraction mediated by antiferromagnetic fluctuations from $J$ overcomes the repulsion from $V_n$. 
For larger values of $J$ and farther away from Van Hove filling, the $d$-wave form factors dominate (see Figs.~\ref{fig:dwave}a,b).
Then, the superconducting ground state is a fully gapped $d+id$ state with $|\mathcal{N}|=2$, which can be seen from the phase of the superconducting state winding two times along the Fermi surface (see Figs.~\ref{fig:dwave}c,d).


Our results highlight twisted hetero-bilayers of TMDs as prime candidates for exotic topological superconducting states in two-dimensional materials. 
They allow --- by moir\'e or substrate engineering --- for an unprecedented level of external control \cite{Kennes2021} and our identification of topological $g+ig$ superconductivity opens up pathways to interrogate this elusive phase of matter in a highly tunable setup. 
Exploiting the unparalleled level of control, these platforms provide the opportunity to scrutinize topological phase transitions using gating which,  as we showed,  
drives the $g+ig$ state into a density wave or a metallic state at either side of the topological superconductor. 
This can also shed light on related questions about topological transitions, e.g., the nodal structure at the transition point from $|\mathcal N|=2$ to $|\mathcal N|=4$, which is intensely debated for Na$_x$CoO$_2$~\cite{PhysRevLett.100.217002,PhysRevLett.111.097001}. 
An intriguing avenue of future research concerns the relevance of (magnetic as well as non-magnetic) disorder and finite size effects onto the $g$-wave superconducting state as moir\'e materials tend to form localized dislocations~\cite{Halbertal2021}. 
From a theoretical angle, the recently developed real-space extension of the unbiased renormalization scheme used here might provide insights into these questions~\cite{PhysRevResearch.3.023180}. 
Another interesting possibility is to investigate if non-local Coulomb interactions can also induce topological triplet superconductivity as discussed for Ref.~\cite{wolf2021triplet}. 

Experimentally, the prediction of $g+ig$ topological superconductivity can be verified using thermal or spin quantum Hall measurements which reveals the four-fold nature of the chiral and topologically protected edge modes. 
Domain walls between $\mathcal N=+4$ and $-4$ configurations must host eight propagating chiral modes~\cite{Volovik1997}. 
Whether these edge modes can be utilized  for 
future quantum information technologies requires additional investigation~\cite{PhysRevLett.109.197001}. 
The option appears particularly intriguing with twisted hetero-bilayers of TMDs being so highly tunable and the energy scales on which the material properties can be altered being so low due to the flat bands. 

\section*{Methods}

We have employed the functional renormalization group method to explore the phase diagram of our model~\cite{RevModPhys.84.299,platt2013functional,Dupuis:2020fhh}. 
Within the FRG, we choose the temperature as the flow parameter and use an approximation that neglects feed-back from the self-energy and three-particle vertices or higher. 
In this approximation, we obtain a renormalization group equation for the two-particle correlation function $\Gamma^{(2p)}$ that describes its evolution upon lowering the temperature. 
In an SU(2)-symmetric system, $\Gamma^{(2p)}$ can be expressed via a (pseudo-)spin-independent coupling function $V$ as $\Gamma^{(2p)}_{s_1s_2s_3s_4}(k_1,k_2,k_3,k_4)=V(k_1,k_2,k_3,k_4)\delta_{s_1s_3}\delta_{s_2s_4}-V(k_1,k_2,k_4,k_3)\delta_{s_1s_4}\delta_{s_2s_3}$, where $s_i$ labels the (pseudo-)spin, and $k_1,k_2$ are incoming and $k_3,k_4$ outgoing momenta. 
Momentum conservation requires $k_1+k_2=k_3+k_4$, so for brevity we will use $V(k_1,k_2,k_3)=V(k_1,k_2,k_3,k_1+k_2-k_3)$ in the following. 

The RG equation for the temperature evolution of $V(k_1,k_2,k_3)$ can then be written as
\begin{align}\label{eq:vertexfrg}
	\frac{d}{d T}V=\tau_\mathrm{pp}+\tau_\mathrm{ph,d}+\tau_\mathrm{ph,cr}\,.
\end{align}
with contributions from the particle-particle, the direct paricle-hole, and the crossed particle-hole channel on the right hand side. They are given by
\begin{align}
	\tau_{\mathrm{pp}}=-\frac{1}{2}\int_{\mathrm{BZ}} d^2k V(k_1,k_2,k)L(k,q_{\mathrm{pp}})V(k,q_{\mathrm{pp}},k_3)\,,\notag
\end{align}
where we used the short hand $\int_{BZ} d^2k=-A^{-1}_{\mathrm{BZ}}\int d^2 k$ and $A_{\mathrm{BZ}}$ is the area of the Brillouin zone.
The particle-hole contributions read 
\begin{align}
	\tau_{\mathrm{ph,d}}=&\frac{1}{2}\int_{\mathrm{BZ}} d^2k [2V(k_1,k,k_3)L(k,q_{\mathrm{d}})V(q_{\mathrm{d}},k_2,k)\notag\\	
&\quad\quad\quad\quad-V(k,k_1,k_3)L(k,q_{\mathrm{d}})V(q_{\mathrm{d}},k_2,k)\notag\\
&\quad\quad\quad\quad-V(k,k_1,k_3)L(k,q_{\mathrm{d}})V(k_2,q_{\mathrm{d}},k)]\,,\notag
\end{align}
and
\begin{align}
	\tau_{\mathrm{ph,cr}}=&-\frac{1}{2}\int_{\mathrm{BZ}} d^2k V(k,k_2,k_3)L(k,q_{\mathrm{cr}})V(k_1,q_{\mathrm{cr}},k)\,.\notag
\end{align}
In these expressions, we introduced $q_{\mathrm{pp}}=-k+k_1+k_2$, $q_{\mathrm{d}}=k+k_1-k_3$, $q_{\mathrm{cr}}=k+k_2-k_3$, and the loop kernel
\begin{align}
L(k,\pm k\!+\!k^\prime)\!=\!\frac{d}{dT}\Big[T\!\sum_{i\omega}G_0(i\omega,k)G_0(\pm i\omega, \pm k\!+\!k^\prime)\Big]\,,\notag
\end{align}
with the free propagator $G_0(i\omega,k)=[i\omega-\xi_k]^{-1}$. 
In these expressions, we have neglected the (external) frequency dependence assuming that the strongest correlations occur for the lowest Matsubara frequencies.

For the numerical implementation, we resolve the momentum dependence in a so-called patching scheme that divides the Fermi surface into $N$ pieces based on equidistant angles and treats the radial dependence for a fixed angle as constant. 
This accurately describes the relevant momentum dependence, which is along the Fermi surface.
In our numerical calculations, we have chosen between $N=48$ and $N=96$ patches, cf. Fig.~\ref{fig:FS}b. 
Our results on the type of instability do not depend on this choice and the quantitative results for critical temperatures vary only mildly with $N$.

The initial condition for $V(k_1,k_2,k_3)$ at high temperatures is given by the Fourier transform of $H_I$ in Eq.~\eqref{eq:model}. 
We set $T_0=\mathrm{max}(\epsilon_{\vec{k}})$ as starting temperature. 
We then calculate the temperature evolution of $V(k_1,k_2,k_3)$ according to Eq.~\eqref{eq:vertexfrg} by solving the integro-differential equation. 
As described above, the development of strong correlations is signaled by a diverging $V(k_1,k_2,k_3)$ at a critical temperature $T_c$. 
Our numerical criterion to detect the divergence is a convex temperature dependence and max$[V(k_1,k_2,k_3)]$ exceeding $30 t_1$.
$T_c$ would be the mean-field critical temperature in an RPA resummation, however, here the estimate is slightly improved due to the inclusion of the coupling between different channels. 
The Fermi liquid is stable within our numerical accuracy if no divergence occurs before $T_l=2\cdot 10^{-4}t_1$ is reached. 
In the cases when correlated states develop, we can read off the type of correlations from the momentum structure of $V(k_1,k_2,k_3)$ at $T_c$. 
Up to an overall constant, this determines the effective interaction close to the instability and directly suggests the order-parameter corresponding to the instability. 
Following this procedure for an extended range of parameters, we obtain the presented phase diagrams.
To extract the form factors of the superconducting instabilities, we diagonalize $V(k,-k,k')$, keep the eigenfunction(s) with the largest eigenvalue and approximate it by lattice harmonics. 

We have used the extracted lattice harmonics to derive the Landau functional \eqref{eq:Landau} from our microscopic model. The decisive prefactor of the term $|\Delta_1^2+\Delta_2^2|^2$ is given by
\begin{align}
    \gamma&=T\sum_{i\omega}\int_{BZ} d^2k \frac{g_1(\vec k)^2 g_2(\vec k)^2}{(i\omega-\xi_{\vec k})^2(i\omega+\xi_{\vec k})^2}\notag \\
    &=\int_{\mathrm{BZ}} d^2k\, g_1(\vec k)^2 g_2(\vec k)^2\frac{1-2n_F(\xi_{\vec k})+2\xi_{\vec k}n_F'(\xi_{\vec k})}{4\xi_{\vec k}^3}
\end{align}
with the Fermi function $n_F$. 
We have calculated $\gamma$ numerically and found it to be positive in the considered range of $\mu$ and $T$. 
As an analytical estimate for $\gamma$, we can approximate the dispersion by $\xi\approx k^2/(2m)-\mu$ with density of states~$\rho_\epsilon$, and the form factors by $g_1=\cos(n\varphi)$, $g_2=\sin(n\varphi)$ with $\varphi=\arctan k_y/k_x$ and $n=4$ for $g+ig$ superconductivity ($n=2$ for $d+id$ and $n=1$ for $p+ip$). 
With this simplification, we obtain
\begin{align}
    \gamma &\approx \int\! d\varphi \cos^2(n\varphi) \sin^2 (n\varphi)\! \int\! d\epsilon \rho_\epsilon\frac{\frac{1}{2}-n_F(\epsilon)+\epsilon n_F'(\epsilon)}{2\epsilon^3}\notag \\
&=\frac{m}{16\pi T^2}\int \frac{\sinh(x)-x}{4x^3(1+\cosh(x))}\approx 0.05\frac{m}{16\pi T^2}\,,
\end{align}
which we can take as a rough estimate for $\gamma$ if $\mu$ is away from the Van Hove energy. 
Right at the Van Hove energy, an additional logarithmic dependence on $T_c$ emerges.\medskip

\paragraph*{Acknowledgements.}--- 
We thank Andrey Chubukov and Abhay Pasupathy for useful discussions. 
MMS acknowledges support by the DFG through SFB 1238 (project C02, project id 277146847).
DMK acknowledges support from the Deutsche  Forschungsgemeinschaft (DFG, German Research Foundation) through RTG 1995, within the Priority Program SPP 2244 ``2DMP'', under Germany’s Excellence Strategy-Cluster of Excellence Matter and Light for Quantum Computing (ML4Q) EXC2004/1 - 390534769, and from the Max Planck-New York City Center for Non-Equilibrium Quantum Phenomena. 
LC was supported by the U.S. Department of Energy (DOE), Office of Basic Energy Sciences, under Contract No. DE- SC0012704.

\paragraph*{Author Contributions.}--- The work was conceived by LC and MMS. MMS computed functional RG data. All authors analyzed and interpreted the results and wrote the manuscript. 
\paragraph*{Competing Interests.}--- The authors declare no competing interests.

\paragraph*{Data availability.}--- Data and simulation codes are available from the corresponding authors upon reasonable request.



\begin{thebibliography}{10}
\expandafter\ifx\csname url\endcsname\relax
  \def\url#1{\texttt{#1}}\fi
\expandafter\ifx\csname urlprefix\endcsname\relax\def\urlprefix{URL }\fi
\providecommand{\bibinfo}[2]{#2}
\providecommand{\eprint}[2][]{\url{#2}}

\bibitem{Cao2018a}
\bibinfo{author}{Cao, Y.} \emph{et~al.}
\newblock \bibinfo{title}{{Unconventional superconductivity in magic-angle
  graphene superlattices}}.
\newblock \emph{\bibinfo{journal}{Nature}} \textbf{\bibinfo{volume}{556}},
  \bibinfo{pages}{43--50} (\bibinfo{year}{2018}).
\newblock \urlprefix\url{http://dx.doi.org/10.1038/nature26160}.

\bibitem{Cao2018}
\bibinfo{author}{Cao, Y.} \emph{et~al.}
\newblock \bibinfo{title}{{Correlated insulator behaviour at half-filling in
  magic-angle graphene superlattices}}.
\newblock \emph{\bibinfo{journal}{Nature}} \textbf{\bibinfo{volume}{556}},
  \bibinfo{pages}{80--84} (\bibinfo{year}{2018}).
\newblock \urlprefix\url{http://dx.doi.org/10.1038/nature26154}.

\bibitem{Yankowitz2019}
\bibinfo{author}{Yankowitz, M.} \emph{et~al.}
\newblock \bibinfo{title}{{Tuning superconductivity in twisted bilayer
  graphene}}.
\newblock \emph{\bibinfo{journal}{Science}} \textbf{\bibinfo{volume}{363}},
  \bibinfo{pages}{1059--1064} (\bibinfo{year}{2019}).
  \newblock \urlprefix\url{http://dx.doi.org/10.1126/science.aav1910}.

\bibitem{Kerelsky2019}
\bibinfo{author}{Kerelsky, A.} \emph{et~al.}
\newblock \bibinfo{title}{Maximized electron interactions at the magic angle in
  twisted bilayer graphene}.
\newblock \emph{\bibinfo{journal}{Nature}} \textbf{\bibinfo{volume}{572}},
  \bibinfo{pages}{95--100} (\bibinfo{year}{2019}).
\newblock \urlprefix\url{https://doi.org/10.1038/s41586-019-1431-9}.

\bibitem{Sharpe605}
\bibinfo{author}{Sharpe, A.~L.} \emph{et~al.}
\newblock \bibinfo{title}{Emergent ferromagnetism near three-quarters filling
  in twisted bilayer graphene}.
\newblock \emph{\bibinfo{journal}{Science}} \textbf{\bibinfo{volume}{365}},
  \bibinfo{pages}{605--608} (\bibinfo{year}{2019}).
  \newblock \urlprefix\url{https://doi.org/10.1126/science.aaw3780}.

\bibitem{lu2019superconductors}
\bibinfo{author}{{Lu}, X.} \emph{et~al.}
\newblock \bibinfo{title}{{Superconductors , orbital magnets and correlated
  states in magic-angle bilayer graphene}}.
\newblock \emph{\bibinfo{journal}{Nature}} \textbf{\bibinfo{volume}{574}},
  \bibinfo{pages}{20--23} (\bibinfo{year}{2019}).
\newblock \urlprefix\url{http://dx.doi.org/10.1038/s41586-019-1695-0}.

\bibitem{Serlin2019}
\bibinfo{author}{Serlin, M.} \emph{et~al.}
\newblock \bibinfo{title}{{Intrinsic quantized anomalous Hall effect in a
  moir{\'{e}} heterostructure}}.
\newblock \emph{\bibinfo{journal}{Science}} \textbf{\bibinfo{volume}{367}},
  \bibinfo{pages}{900--903} (\bibinfo{year}{2020}).
  \newblock \urlprefix\url{http://dx.doi.org/10.1126/science.aay5533}.

\bibitem{Kennes2021}
\bibinfo{author}{Kennes, D.~M.} \emph{et~al.}
\newblock \bibinfo{title}{Moir{\'e} heterostructures as a condensed-matter
  quantum simulator}.
\newblock \emph{\bibinfo{journal}{Nature Physics}}
  \textbf{\bibinfo{volume}{17}}, \bibinfo{pages}{155--163}
  (\bibinfo{year}{2021}).
\newblock \urlprefix\url{https://doi.org/10.1038/s41567-020-01154-3}.

\bibitem{Liu2019}
\bibinfo{author}{Liu, X.} \emph{et~al.}
\newblock \bibinfo{title}{Tunable spin-polarized correlated states in twisted
  double bilayer graphene}.
\newblock \emph{\bibinfo{journal}{Nature}} \textbf{\bibinfo{volume}{583}},
  \bibinfo{pages}{221--225} (\bibinfo{year}{2020}).
\newblock \urlprefix\url{https://doi.org/10.1038/s41586-020-2458-7}.

\bibitem{Cao2019}
\bibinfo{author}{Cao, Y.} \emph{et~al.}
\newblock \bibinfo{title}{Tunable correlated states and spin-polarized phases
  in twisted bilayer--bilayer graphene}.
\newblock \emph{\bibinfo{journal}{Nature}} \textbf{\bibinfo{volume}{583}},
  \bibinfo{pages}{215--220} (\bibinfo{year}{2020}).
\newblock \urlprefix\url{https://doi.org/10.1038/s41586-020-2260-6}.

\bibitem{Shen2019}
\bibinfo{author}{Shen, C.} \emph{et~al.}
\newblock \bibinfo{title}{Correlated states in twisted double bilayer
  graphene}.
\newblock \emph{\bibinfo{journal}{Nature Physics}}
  \textbf{\bibinfo{volume}{16}}, \bibinfo{pages}{520--525}
  (\bibinfo{year}{2020}).
\newblock \urlprefix\url{https://doi.org/10.1038/s41567-020-0825-9}.

\bibitem{Chen2019a}
\bibinfo{author}{Chen, G.} \emph{et~al.}
\newblock \bibinfo{title}{{Evidence of a gate-tunable Mott insulator in a
  trilayer graphene moir{\'{e}} superlattice}}.
\newblock \emph{\bibinfo{journal}{Nature Physics}}
  \textbf{\bibinfo{volume}{15}}, \bibinfo{pages}{237--241}
  (\bibinfo{year}{2019}).
\newblock \urlprefix\url{http://dx.doi.org/10.1038/s41567-018-0387-2}.

\bibitem{Chen2019}
\bibinfo{author}{Chen, G.} \emph{et~al.}
\newblock \bibinfo{title}{Signatures of tunable superconductivity in a trilayer
  graphene moir{\'e} superlattice}.
\newblock \emph{\bibinfo{journal}{Nature}} \textbf{\bibinfo{volume}{572}},
  \bibinfo{pages}{215--219} (\bibinfo{year}{2019}).
\newblock \urlprefix\url{https://doi.org/10.1038/s41586-019-1393-y}.

\bibitem{chen2019tunable}
\bibinfo{author}{Chen, G.} \emph{et~al.}
\newblock \bibinfo{title}{Tunable correlated chern insulator and ferromagnetism
  in a moir{\'e} superlattice}.
\newblock \emph{\bibinfo{journal}{Nature}} \textbf{\bibinfo{volume}{579}},
  \bibinfo{pages}{56--61} (\bibinfo{year}{2020}).
\newblock \urlprefix\url{https://doi.org/10.1038/s41586-020-2049-7}.

\bibitem{TutucBi}
\bibinfo{author}{Burg, G.~W.} \emph{et~al.}
\newblock \bibinfo{title}{Correlated insulating states in twisted double
  bilayer graphene}.
\newblock \emph{\bibinfo{journal}{Phys. Rev. Lett.}}
  \textbf{\bibinfo{volume}{123}}, \bibinfo{pages}{197702}
  (\bibinfo{year}{2019}).
\newblock
  \urlprefix\url{https://link.aps.org/doi/10.1103/PhysRevLett.123.197702}.

\bibitem{rubioverdu2020universal}
\bibinfo{author}{Rubio-Verdú, C.} \emph{et~al.}
\newblock \bibinfo{title}{Universal moir\'e nematic phase in twisted graphitic
  systems}.
\newblock \emph{\bibinfo{journal}{arXiv:2009.11645}}  (\bibinfo{year}{2020}).
\newblock \urlprefix\url{https://arxiv.org/abs/2009.11645}.

\bibitem{Xian2019BN}
\bibinfo{author}{Xian, L.}, \bibinfo{author}{Kennes, D.~M.},
  \bibinfo{author}{Tancogne-Dejean, N.}, \bibinfo{author}{Altarelli, M.} \&
  \bibinfo{author}{Rubio, A.}
\newblock \bibinfo{title}{Multiflat bands and strong correlations in twisted
  bilayer boron nitride: Doping-induced correlated insulator and
  superconductor}.
\newblock \emph{\bibinfo{journal}{Nano Letters}} \textbf{\bibinfo{volume}{19}},
  \bibinfo{pages}{4934--4940} (\bibinfo{year}{2019}).
\newblock \urlprefix\url{https://doi.org/10.1021/acs.nanolett.9b00986}.

\bibitem{Wu2018}
\bibinfo{author}{Wu, F.}, \bibinfo{author}{Lovorn, T.}, \bibinfo{author}{Tutuc,
  E.} \& \bibinfo{author}{Macdonald, A.~H.}
\newblock \bibinfo{title}{{Hubbard Model Physics in Transition Metal
  Dichalcogenide Moir{\'{e}} Bands}}.
\newblock \emph{\bibinfo{journal}{Phys. Rev. Lett.}}
  \textbf{\bibinfo{volume}{121}}, \bibinfo{pages}{26402}
  (\bibinfo{year}{2018}).
\newblock \urlprefix\url{https://doi.org/10.1103/PhysRevLett.121.026402}.

\bibitem{Wu2019}
\bibinfo{author}{Wu, F.}, \bibinfo{author}{Lovorn, T.}, \bibinfo{author}{Tutuc,
  E.}, \bibinfo{author}{Martin, I.} \& \bibinfo{author}{Macdonald, A.~H.}
\newblock \bibinfo{title}{{Topological Insulators in Twisted Transition Metal
  Dichalcogenide Homobilayers}}.
\newblock \emph{\bibinfo{journal}{Phys. Rev. Lett.}}
  \textbf{\bibinfo{volume}{122}}, \bibinfo{pages}{86402}
  (\bibinfo{year}{2019}).
\newblock \urlprefix\url{https://doi.org/10.1103/PhysRevLett.122.086402}.

\bibitem{Naik2018}
\bibinfo{author}{Naik, M.~H.} \& \bibinfo{author}{Jain, M.}
\newblock \bibinfo{title}{{Ultraflatbands and Shear Solitons in Moir{\'{e}}
  Patterns of Twisted Bilayer Transition Metal Dichalcogenides}}.
\newblock \emph{\bibinfo{journal}{Phys. Rev. Lett.}}
  \textbf{\bibinfo{volume}{121}}, \bibinfo{pages}{266401}
  (\bibinfo{year}{2018}).
\newblock \urlprefix\url{https://doi.org/10.1103/PhysRevLett.121.266401}.

\bibitem{Ruiz-Tijerina2019}
\bibinfo{author}{Ruiz-Tijerina, D.~A.} \& \bibinfo{author}{Fal'Ko, V.~I.}
\newblock \bibinfo{title}{{Interlayer hybridization and moir{\'{e}}
  superlattice minibands for electrons and excitons in heterobilayers of
  transition-metal dichalcogenides}}.
\newblock \emph{\bibinfo{journal}{Phys. Rev. B}} \textbf{\bibinfo{volume}{99}},
  \bibinfo{pages}{30--32} (\bibinfo{year}{2019}).
 \newblock \urlprefix\url{https://link.aps.org/doi/10.1103/PhysRevB.99.125424}.

\bibitem{Schrade2019}
\bibinfo{author}{Schrade, C.} \& \bibinfo{author}{Fu, L.}
\newblock \bibinfo{title}{{Spin-valley density wave in moir{\'{e}} materials}}.
\newblock \emph{\bibinfo{journal}{Phys. Rev. B}}
  \textbf{\bibinfo{volume}{100}}, \bibinfo{pages}{035413}
  (\bibinfo{year}{2019}).
\newblock \urlprefix\url{https://link.aps.org/doi/10.1103/PhysRevB.100.035413}.

\bibitem{Wang2020WSe2}
\bibinfo{author}{Wang, L.} \emph{et~al.}
\newblock \bibinfo{title}{Correlated electronic phases in twisted bilayer
  transition metal dichalcogenides}.
\newblock \emph{\bibinfo{journal}{Nature Materials}}
  \textbf{\bibinfo{volume}{19}}, \bibinfo{pages}{861--866}
  (\bibinfo{year}{2020}).
\newblock \urlprefix\url{https://doi.org/10.1038/s41563-020-0708-6}.

\bibitem{zhou2021quantum}
\bibinfo{author}{Zhou, Y.}, \bibinfo{author}{Sheng, D.} \&
  \bibinfo{author}{Kim, E.-A.}
\newblock \bibinfo{title}{Quantum phases of transition metal dichalcogenide
  moir\'e systems}.
\newblock \emph{\bibinfo{journal}{arXiv:2105.07008}}
  (\bibinfo{year}{2021}).
  \newblock \urlprefix\url{https://arxiv.org/abs/2105.07008}.

\bibitem{Jin2019ex}
\bibinfo{author}{Jin, C.} \emph{et~al.}
\newblock \bibinfo{title}{{Observation of moir{\'e} excitons in WSe$_2$/WS$_2$
  heterostructure superlattices}}.
\newblock \emph{\bibinfo{journal}{Nature}} \textbf{\bibinfo{volume}{567}},
  \bibinfo{pages}{76--80} (\bibinfo{year}{2019}).
  \newblock \urlprefix\url{https://doi.org/10.1038/s41586-019-0976-y}.

\bibitem{Wang2019ex}
\bibinfo{author}{Wang, Z.} \emph{et~al.}
\newblock \bibinfo{title}{Evidence of high-temperature exciton condensation in
  two-dimensional atomic double layers}.
\newblock \emph{\bibinfo{journal}{Nature}} \textbf{\bibinfo{volume}{574}},
  \bibinfo{pages}{76--80} (\bibinfo{year}{2019}).
   \newblock \urlprefix\url{https://doi.org/10.1038/s41586-019-1591-7}.

\bibitem{shimazaki2020stronglyex}
\bibinfo{author}{Shimazaki, Y.} \emph{et~al.}
\newblock \bibinfo{title}{Strongly correlated electrons and hybrid excitons in
  a moir{\'e} heterostructure}.
\newblock \emph{\bibinfo{journal}{Nature}} \textbf{\bibinfo{volume}{580}},
  \bibinfo{pages}{472--477} (\bibinfo{year}{2020}).
 \newblock \urlprefix\url{https://doi.org/10.1038/s41586-020-2191-2}.

\bibitem{tang2019wse2}
\bibinfo{author}{Tang, Y.} \emph{et~al.}
\newblock \bibinfo{title}{{WSe${}_2$/WS${}_2$ moir\'e superlattices: a new
  Hubbard model simulator}}.
\newblock \emph{\bibinfo{journal}{arXiv:1910.08673}}
  (\bibinfo{year}{2019}).
   \newblock \urlprefix\url{https://arxiv.org/abs/1910.08673}.

\bibitem{Regan2020}
\bibinfo{author}{Regan, E.~C.} \emph{et~al.}
\newblock \bibinfo{title}{{Mott and generalized Wigner crystal states in
  WSe2/WS2 moir{\'e} superlattices}}.
\newblock \emph{\bibinfo{journal}{Nature}} \textbf{\bibinfo{volume}{579}},
  \bibinfo{pages}{359--363} (\bibinfo{year}{2020}).
\newblock \urlprefix\url{https://doi.org/10.1038/s41586-020-2092-4}.

\bibitem{Jin2021}
\bibinfo{author}{Jin, C.} \emph{et~al.}
\newblock \bibinfo{title}{{Stripe phases in WSe2/WS2 moir{\'e} superlattices}}.
\newblock \emph{\bibinfo{journal}{Nature Materials}}
  \textbf{\bibinfo{volume}{20}}, \bibinfo{pages}{940--944}
  (\bibinfo{year}{2021}).
\newblock \urlprefix\url{https://doi.org/10.1038/s41563-021-00959-8}.

\bibitem{li2021quantum}
\bibinfo{author}{Li, T.} \emph{et~al.}
\newblock \bibinfo{title}{{Quantum anomalous Hall effect from intertwined
  moir\'e bands}}.
  \newblock \emph{\bibinfo{journal}{arXiv:2107.01796}}
  (\bibinfo{year}{2021}).
\newblock \urlprefix\url{https://arxiv.org/abs/2107.01796}.

\bibitem{RevModPhys.75.473}
\bibinfo{author}{Damascelli, A.}, \bibinfo{author}{Hussain, Z.} \&
  \bibinfo{author}{Shen, Z.-X.}
\newblock \bibinfo{title}{Angle-resolved photoemission studies of the cuprate
  superconductors}.
\newblock \emph{\bibinfo{journal}{Rev. Mod. Phys.}}
  \textbf{\bibinfo{volume}{75}}, \bibinfo{pages}{473--541}
  (\bibinfo{year}{2003}).
\newblock \urlprefix\url{https://link.aps.org/doi/10.1103/RevModPhys.75.473}.

\bibitem{RevModPhys.79.353}
\bibinfo{author}{Fischer, O.}, \bibinfo{author}{Kugler, M.},
  \bibinfo{author}{Maggio-Aprile, I.}, \bibinfo{author}{Berthod, C.} \&
  \bibinfo{author}{Renner, C.}
\newblock \bibinfo{title}{Scanning tunneling spectroscopy of high-temperature
  superconductors}.
\newblock \emph{\bibinfo{journal}{Rev. Mod. Phys.}}
  \textbf{\bibinfo{volume}{79}}, \bibinfo{pages}{353--419}
  (\bibinfo{year}{2007}).
\newblock \urlprefix\url{https://link.aps.org/doi/10.1103/RevModPhys.79.353}.

\bibitem{Bardeen1209}
\bibinfo{author}{Bardeen, J.}
\newblock \bibinfo{title}{Electron-phonon interactions and superconductivity}.
\newblock \emph{\bibinfo{journal}{Science}} \textbf{\bibinfo{volume}{181}},
  \bibinfo{pages}{1209--1214} (\bibinfo{year}{1973}).
\newblock \urlprefix\url{https://science.sciencemag.org/content/181/4106/1209}.

\bibitem{RevModPhys.80.1083}
\bibinfo{author}{Nayak, C.}, \bibinfo{author}{Simon, S.~H.},
  \bibinfo{author}{Stern, A.}, \bibinfo{author}{Freedman, M.} \&
  \bibinfo{author}{Das~Sarma, S.}
\newblock \bibinfo{title}{Non-abelian anyons and topological quantum
  computation}.
\newblock \emph{\bibinfo{journal}{Rev. Mod. Phys.}}
  \textbf{\bibinfo{volume}{80}}, \bibinfo{pages}{1083--1159}
  (\bibinfo{year}{2008}).
\newblock \urlprefix\url{https://link.aps.org/doi/10.1103/RevModPhys.80.1083}.

\bibitem{UPTRMP}
\bibinfo{author}{Joynt, R.} \& \bibinfo{author}{Taillefer, L.}
\newblock \bibinfo{title}{{The superconducting phases of
  ${\mathrm{UPt}}_{3}$}}.
\newblock \emph{\bibinfo{journal}{Rev. Mod. Phys.}}
  \textbf{\bibinfo{volume}{74}}, \bibinfo{pages}{235--294}
  (\bibinfo{year}{2002}).
\newblock \urlprefix\url{https://link.aps.org/doi/10.1103/RevModPhys.74.235}.

\bibitem{UPt3}
\bibinfo{author}{Avers, K.~E.} \emph{et~al.}
\newblock \bibinfo{title}{{Broken time-reversal symmetry in the topological
  superconductor UPt${}_3$}}.
\newblock \emph{\bibinfo{journal}{Nature Physics}}
  \textbf{\bibinfo{volume}{16}}, \bibinfo{pages}{531--535}
  (\bibinfo{year}{2020}).
\newblock \urlprefix\url{https://doi.org/10.1038/s41567-020-0822-z}.

\bibitem{UTe2}
\bibinfo{author}{Jiao, L.} \emph{et~al.}
\newblock \bibinfo{title}{{Chiral superconductivity in heavy-fermion metal
  UTe${}_2$}}.
\newblock \emph{\bibinfo{journal}{Nature}} \textbf{\bibinfo{volume}{579}},
  \bibinfo{pages}{523--527} (\bibinfo{year}{2020}).
\newblock \urlprefix\url{https://doi.org/10.1038/s41586-020-2122-2}.

\bibitem{Zhang182}
\bibinfo{author}{Zhang, P.} \emph{et~al.}
\newblock \bibinfo{title}{{Observation of topological superconductivity on the
  surface of an iron-based superconductor}}.
\newblock \emph{\bibinfo{journal}{Science}} \textbf{\bibinfo{volume}{360}},
  \bibinfo{pages}{182--186} (\bibinfo{year}{2018}).
\newblock \urlprefix\url{https://science.sciencemag.org/content/360/6385/182}.

\bibitem{Li2021}
\bibinfo{author}{Li, Y.} \emph{et~al.}
\newblock \bibinfo{title}{{Electronic properties of the bulk and surface states
  of Fe$_{1+y}$Te$_{1-x}$Se$_x$}}.
\newblock \emph{\bibinfo{journal}{Nature Materials}}  (\bibinfo{year}{2021}).
\newblock \urlprefix\url{https://doi.org/10.1038/s41563-021-00984-7}.

\bibitem{Lian20}
\bibinfo{author}{Lian, B.}, \bibinfo{author}{Liu, Z.}, \bibinfo{author}{Zhang,
  Y.} \& \bibinfo{author}{Wang, J.}
\newblock \bibinfo{title}{Flat chern band from twisted bilayer
  {MnBi$_{2}$Te$_{4}$}}.
\newblock \emph{\bibinfo{journal}{Phys. Rev. Lett.}}
  \textbf{\bibinfo{volume}{124}}, \bibinfo{pages}{126402}
  (\bibinfo{year}{2020}).
\newblock
  \urlprefix\url{https://link.aps.org/doi/10.1103/PhysRevLett.124.126402}.

\bibitem{KennesGeSe}
\bibinfo{author}{Kennes, D.~M.}, \bibinfo{author}{Xian, L.},
  \bibinfo{author}{Claassen, M.} \& \bibinfo{author}{Rubio, A.}
\newblock \bibinfo{title}{One-dimensional flat bands in twisted bilayer
  germanium selenide}.
\newblock \emph{\bibinfo{journal}{Nature Communications}}
  \textbf{\bibinfo{volume}{11}}, \bibinfo{pages}{1124} (\bibinfo{year}{2020}).
\newblock \urlprefix\url{https://doi.org/10.1038/s41467-020-14947-0}.

\bibitem{duran2021moire}
\bibinfo{author}{Morales-Dur\'an, N.}, \bibinfo{author}{Hu, N.~C.},
  \bibinfo{author}{Potasz, P.} \& \bibinfo{author}{MacDonald, A.~H.}
\newblock \bibinfo{title}{{Non-local interactions in moir\'e Hubbard systems}}.
\newblock \emph{\bibinfo{journal}{arXiv:2108.03313}}
  (\bibinfo{year}{2021}).
  \newblock \urlprefix\url{https://arxiv.org/abs/2108.03313}.

\bibitem{RevModPhys.84.299}
\bibinfo{author}{Metzner, W.}, \bibinfo{author}{Salmhofer, M.},
  \bibinfo{author}{Honerkamp, C.}, \bibinfo{author}{Meden, V.} \&
  \bibinfo{author}{Sch\"onhammer, K.}
\newblock \bibinfo{title}{{Functional renormalization group approach to
  correlated fermion systems}}.
\newblock \emph{\bibinfo{journal}{Rev. Mod. Phys.}}
  \textbf{\bibinfo{volume}{84}}, \bibinfo{pages}{299--352}
  (\bibinfo{year}{2012}).
\newblock \urlprefix\url{https://link.aps.org/doi/10.1103/RevModPhys.84.299}.

\bibitem{PhysRevLett.108.227204}
\bibinfo{author}{Nandkishore, R.}, \bibinfo{author}{Chern, G.-W.} \&
  \bibinfo{author}{Chubukov, A.~V.}
\newblock \bibinfo{title}{{Itinerant Half-Metal Spin-Density-Wave State on the
  Hexagonal Lattice}}.
\newblock \emph{\bibinfo{journal}{Phys. Rev. Lett.}}
  \textbf{\bibinfo{volume}{108}}, \bibinfo{pages}{227204}
  (\bibinfo{year}{2012}).
\newblock
  \urlprefix\url{https://link.aps.org/doi/10.1103/PhysRevLett.108.227204}.

\bibitem{PhysRevLett.101.156402}
\bibinfo{author}{Martin, I.} \& \bibinfo{author}{Batista, C.~D.}
\newblock \bibinfo{title}{{Itinerant Electron-Driven Chiral Magnetic Ordering
  and Spontaneous Quantum Hall Effect in Triangular Lattice Models}}.
\newblock \emph{\bibinfo{journal}{Phys. Rev. Lett.}}
  \textbf{\bibinfo{volume}{101}}, \bibinfo{pages}{156402}
  (\bibinfo{year}{2008}).
\newblock
  \urlprefix\url{https://link.aps.org/doi/10.1103/PhysRevLett.101.156402}.

\bibitem{PhysRevB.89.144501}
\bibinfo{author}{Nandkishore, R.}, \bibinfo{author}{Thomale, R.} \&
  \bibinfo{author}{Chubukov, A.~V.}
\newblock \bibinfo{title}{Superconductivity from weak repulsion in hexagonal
  lattice systems}.
\newblock \emph{\bibinfo{journal}{Phys. Rev. B}} \textbf{\bibinfo{volume}{89}},
  \bibinfo{pages}{144501} (\bibinfo{year}{2014}).
\newblock \urlprefix\url{https://link.aps.org/doi/10.1103/PhysRevB.89.144501}.

\bibitem{PhysRevB.68.104510}
\bibinfo{author}{Honerkamp, C.}
\newblock \bibinfo{title}{Instabilities of interacting electrons on the
  triangular lattice}.
\newblock \emph{\bibinfo{journal}{Phys. Rev. B}} \textbf{\bibinfo{volume}{68}},
  \bibinfo{pages}{104510} (\bibinfo{year}{2003}).
\newblock \urlprefix\url{https://link.aps.org/doi/10.1103/PhysRevB.68.104510}.

\bibitem{note1}
\bibinfo{note}{Note that the number of nodes on the Fermi surface depends on
  the chemical potential. It is four if the Fermi surface is closer to $\Gamma$
  (where we do not find a superconducting instability), and eight near Van Hove
  energy (where we do find one)}.

\bibitem{RevModPhys.63.239}
\bibinfo{author}{Sigrist, M.} \& \bibinfo{author}{Ueda, K.}
\newblock \bibinfo{title}{{Phenomenological theory of unconventional
  superconductivity}}.
\newblock \emph{\bibinfo{journal}{Rev. Mod. Phys.}}
  \textbf{\bibinfo{volume}{63}}, \bibinfo{pages}{239--311}
  (\bibinfo{year}{1991}).
\newblock \urlprefix\url{https://link.aps.org/doi/10.1103/RevModPhys.63.239}.

\bibitem{Volovik1997}
\bibinfo{author}{Volovik, G.~E.}
\newblock \bibinfo{title}{{On edge states in superconductors with time
  inversion symmetry breaking}}.
\newblock \emph{\bibinfo{journal}{Journal of Experimental and Theoretical
  Physics Letters}} \textbf{\bibinfo{volume}{66}}, \bibinfo{pages}{522--527}
  (\bibinfo{year}{1997}).
\newblock \urlprefix\url{https://doi.org/10.1134/1.567563}.

\bibitem{PhysRevB.61.10267}
\bibinfo{author}{Read, N.} \& \bibinfo{author}{Green, D.}
\newblock \bibinfo{title}{Paired states of fermions in two dimensions with
  breaking of parity and time-reversal symmetries and the fractional quantum
  Hall effect}.
\newblock \emph{\bibinfo{journal}{Phys. Rev. B}} \textbf{\bibinfo{volume}{61}},
  \bibinfo{pages}{10267--10297} (\bibinfo{year}{2000}).
\newblock \urlprefix\url{https://link.aps.org/doi/10.1103/PhysRevB.61.10267}.

\bibitem{Black_Schaffer_2014}
\bibinfo{author}{Black-Schaffer, A.~M.} \& \bibinfo{author}{Honerkamp, C.}
\newblock \bibinfo{title}{Chiral $d$-wave superconductivity in doped graphene}.
\newblock \emph{\bibinfo{journal}{Journal of Physics: Condensed Matter}}
  \textbf{\bibinfo{volume}{26}}, \bibinfo{pages}{423201}
  (\bibinfo{year}{2014}).
\newblock \urlprefix\url{https://doi.org/10.1088/0953-8984/26/42/423201}.

\bibitem{PhysRevB.60.4245}
\bibinfo{author}{Senthil, T.}, \bibinfo{author}{Marston, J.~B.} \&
  \bibinfo{author}{Fisher, M. P.~A.}
\newblock \bibinfo{title}{{Spin quantum Hall effect in unconventional
  superconductors}}.
\newblock \emph{\bibinfo{journal}{Phys. Rev. B}} \textbf{\bibinfo{volume}{60}},
  \bibinfo{pages}{4245--4254} (\bibinfo{year}{1999}).
\newblock \urlprefix\url{https://link.aps.org/doi/10.1103/PhysRevB.60.4245}.

\bibitem{PhysRevB.68.214503}
\bibinfo{author}{Horovitz, B.} \& \bibinfo{author}{Golub, A.}
\newblock \bibinfo{title}{{Superconductors with broken time-reversal symmetry:
  Spontaneous magnetization and quantum Hall effects}}.
\newblock \emph{\bibinfo{journal}{Phys. Rev. B}} \textbf{\bibinfo{volume}{68}},
  \bibinfo{pages}{214503} (\bibinfo{year}{2003}).
\newblock \urlprefix\url{https://link.aps.org/doi/10.1103/PhysRevB.68.214503}.

\bibitem{PhysRevLett.100.217002}
\bibinfo{author}{Zhou, S.} \& \bibinfo{author}{Wang, Z.}
\newblock \bibinfo{title}{{Nodal $d+id$ Pairing and Topological Phases on the
  Triangular Lattice of
  ${\mathrm{Na}}_{x}{\mathrm{CoO}}_{2}\ifmmode\cdot\else\textperiodcentered\fi{}y{\mathrm{H}}_{2}\mathrm{O}$:
  Evidence for an Unconventional Superconducting State}}.
\newblock \emph{\bibinfo{journal}{Phys. Rev. Lett.}}
  \textbf{\bibinfo{volume}{100}}, \bibinfo{pages}{217002}
  (\bibinfo{year}{2008}).
\newblock
  \urlprefix\url{https://link.aps.org/doi/10.1103/PhysRevLett.100.217002}.

\bibitem{PhysRevLett.111.097001}
\bibinfo{author}{Kiesel, M.~L.}, \bibinfo{author}{Platt, C.},
  \bibinfo{author}{Hanke, W.} \& \bibinfo{author}{Thomale, R.}
\newblock \bibinfo{title}{Model evidence of an anisotropic chiral
  $d\mathbf{+}id$-wave pairing state for the water-intercalated
  ${\mathrm{Na}}_{x}{\mathrm{CoO}}_{2}\ifmmode\cdot\else\textperiodcentered\fi{}y{\mathbf{H}}_{2}\mathbf{O}$
  superconductor}.
\newblock \emph{\bibinfo{journal}{Phys. Rev. Lett.}}
  \textbf{\bibinfo{volume}{111}}, \bibinfo{pages}{097001}
  (\bibinfo{year}{2013}).
\newblock
  \urlprefix\url{https://link.aps.org/doi/10.1103/PhysRevLett.111.097001}.

\bibitem{Halbertal2021}
\bibinfo{author}{Halbertal, D.} \emph{et~al.}
\newblock \bibinfo{title}{Moir{\'e} metrology of energy landscapes in van der
  waals heterostructures}.
\newblock \emph{\bibinfo{journal}{Nature Communications}}
  \textbf{\bibinfo{volume}{12}}, \bibinfo{pages}{242} (\bibinfo{year}{2021}).
\newblock \urlprefix\url{https://doi.org/10.1038/s41467-020-20428-1}.

\bibitem{PhysRevResearch.3.023180}
\bibinfo{author}{Hauck, J.~B.}, \bibinfo{author}{Honerkamp, C.},
  \bibinfo{author}{Achilles, S.} \& \bibinfo{author}{Kennes, D.~M.}
\newblock \bibinfo{title}{Electronic instabilities in Penrose quasicrystals:
  Competition, coexistence, and collaboration of order}.
\newblock \emph{\bibinfo{journal}{Phys. Rev. Research}}
  \textbf{\bibinfo{volume}{3}}, \bibinfo{pages}{023180} (\bibinfo{year}{2021}).
\newblock
  \urlprefix\url{https://link.aps.org/doi/10.1103/PhysRevResearch.3.023180}.

\bibitem{wolf2021triplet}
\bibinfo{author}{Wolf, S.}, \bibinfo{author}{Di~Sante, D.},
  \bibinfo{author}{Schwemmer, T.}, \bibinfo{author}{Thomale, R.} \&
  \bibinfo{author}{Rachel, S.}
\newblock \bibinfo{title}{Triplet superconductivity from non-local Coulomb
  repulsion in Sn/Si (111)}.
\newblock \emph{\bibinfo{journal}{arXiv:2107.03482}}
  (\bibinfo{year}{2021}).
  \newblock
  \urlprefix\url{https://arxiv.org/abs/2107.03482}.

\bibitem{PhysRevLett.109.197001}
\bibinfo{author}{Black-Schaffer, A.~M.}
\newblock \bibinfo{title}{{Edge Properties and Majorana Fermions in the
  Proposed Chiral $d$-Wave Superconducting State of Doped Graphene}}.
\newblock \emph{\bibinfo{journal}{Phys. Rev. Lett.}}
  \textbf{\bibinfo{volume}{109}}, \bibinfo{pages}{197001}
  (\bibinfo{year}{2012}).
\newblock
  \urlprefix\url{https://link.aps.org/doi/10.1103/PhysRevLett.109.197001}.

\bibitem{platt2013functional}
\bibinfo{author}{Platt, C.}, \bibinfo{author}{Hanke, W.} \&
  \bibinfo{author}{Thomale, R.}
\newblock \bibinfo{title}{Functional renormalization group for multi-orbital
  fermi surface instabilities}.
\newblock \emph{\bibinfo{journal}{Advances in Physics}}
  \textbf{\bibinfo{volume}{62}}, \bibinfo{pages}{453--562}
  (\bibinfo{year}{2013}).
  \newblock
  \urlprefix\url{https://doi.org/10.1080/00018732.2013.862020}.

\bibitem{Dupuis:2020fhh}
\bibinfo{author}{Dupuis, N.} \emph{et~al.}
\newblock \bibinfo{title}{{The nonperturbative functional renormalization group
  and its applications}}.
  \newblock \emph{\bibinfo{journal}{Physics Reports}}
  \textbf{\bibinfo{volume}{910}}, \bibinfo{pages}{1--114}
  (\bibinfo{year}{2021}).
  \newblock
  \urlprefix\url{https://doi.org/10.1016/j.physrep.2021.01.001}.

\end{thebibliography}
\end{document}